\documentclass[%
 aip,
 amsmath,amssymb,
 reprint,%
]{revtex4-1}

\usepackage{graphicx}
\usepackage{dcolumn}
\usepackage{bm}

\usepackage[utf8]{inputenc}
\usepackage[T1]{fontenc}
\usepackage{mathptmx}
\usepackage{hyperref}
\begin{document}

\preprint{AIP/123-QED}

\title[Aging transition in the absence of  inactive oscillators ]{Aging transition in the absence of inactive oscillators}

\author{K. Sathiyadevi}
\affiliation{ 
	Centre for Nonlinear Science \& Engineering, School of Electrical \& Electronics Engineering, SASTRA Deemed University, Thanjavur - 613 401, Tamil Nadu, India.
}
\author{I. Gowthaman}%
\affiliation{ 
Centre for Nonlinear Science \& Engineering, School of Electrical \& Electronics Engineering, SASTRA Deemed University, Thanjavur - 613 401, Tamil Nadu, India.
}%
\author{D. V. Senthilkumar}%
\email{skumarusnld@gmail.com}
\affiliation{ School of Physics,
	Indian Institute of Science Education and Research,
	Thiruvananthapuram - 695551,
	Kerala, India.
}%

\author{V. K. Chandrasekar}
 \email{chandru25nld@gmail.com}
\affiliation{ 
	Centre for Nonlinear Science \& Engineering, School of Electrical \& Electronics Engineering, SASTRA Deemed University, Thanjavur - 613 401, Tamil Nadu, India.
}%

\date{\today}

\begin{abstract}
The role of counter-rotating oscillators in an ensemble of coexisting co- and counter-rotating oscillators is examined by increasing the proportion of the latter. The phenomenon of aging transition was identified at a critical value of the ratio of  the counter-rotating oscillators, which was otherwise realized only by increasing the number of inactive oscillators to  a large extent.  The effect of  the mean-field feedback strength in the symmetry preserving coupling is also explored. The parameter space of aging transition was increased abruptly even for a feeble decrease in the feedback strength and subsequently aging transition was observed at a critical value of the feedback strength surprisingly without any counter-rotating oscillators.  Further, the study was extended to symmetry breaking coupling using conjugate variables and it was observed that the symmetry breaking coupling can facilitating the onset of aging transition even in the absence of counter-rotating oscillators and for the unit value of  the feedback strength. In general, the parameter space of aging transition was found to increase by increasing the frequency of oscillators and by increasing the proportion of the counter-rotating oscillators in both the symmetry preserving and symmetry breaking couplings.  Further, the transition from oscillatory to aging transition occurs via a Hopf bifurcation, while the transition from aging transition to oscillation death state emerges via Pitchfork bifurcation.  Analytical expressions for the critical ratio of the counter-rotating oscillators are deduced to find the stable boundaries of the aging transition.    
\end{abstract}

\maketitle

\begin{quotation}
Aging is a kind of deterioration which occurs in diverse complex systems. It is evident from our daily life that living organisms (and its efficiency) degrade as it becomes older. For instance, Alzheimer's disease is an example of a cause of failure of neurons due to the aging process.  In this context, the phenomenon of aging transition was reported in an ensemble of oscillators by increasing the proportion of inactive oscillator \cite{dai1,dai2,dai3,dai4,bhumi,srilena}. In the present work, we show that even  an appropriate proportion of counter-rotating oscillators in an ensemble of coexisting co- and counter-rotating oscillator is capable of inducing the phenomenon of aging transition. Further, we find that either the mean-field feedback or the symmetry breaking coupling alone can facilitate the onset of the aging transition. 

\end{quotation}

\section{Introduction}
Complex systems are abundant, omnipresent in nature and  are rarely isolated.   Coupled nonlinear oscillators serve as an excellent framework to  unravel   distinct collective patterns observed in such complex systems, which include chemical \cite{kuro_book,chem1} and biological systems \cite{neu1,neu2}, and so on.   The network architecture and the strength of the interaction  of the coupled systems  facilitate the onset of  various collective behaviors such as  synchronization \cite{syn1,syn2},  chimera \cite{chim_rev1,chim1,neu2,chim_sch1,chim_sch2,sharika,sb1}, clustering \cite{clus1,clus2}, and distinct oscillation quenching states\cite{od1,od2,od3,ad1,ad2,ad3},   having strong resemblance with many  natural processes.  The phenomenon of  aging transition is one  such a phenomenon of   serious concern   in the complex networks, degrading its dynamical activity.  In   neural networks, de-actuation of a single node may cause series concern in the entire network, which  cascades continuous degradation of neighboring nodes \cite{brain1,brain2}.  A similar issue has also been reported in the power grids \cite{power1}.    Initially, the phenomenon of aging transition was reported by increasing the number of  inactive oscillators from the active group of oscillators in a globally coupled network.  Originally, the effect of aging behavior was analyzed through globally and diffusively coupled oscillators   \cite{dai1,dai2,dai3,dai4}. Further,  the emergence of desynchronization horn  was identified  for sufficiently strong nonisochronicity where the active oscillators desynchronize resulting in local clustering \cite{dai2}.  The phenomenon of   aging transition was also analyzed by different groups  probing  various aspects of the network \cite{dai3,bhumi,srilena,sen}.

On the other hand, coexisting co- and counter-rotating dynamical activity can be found in various natural systems such as     fluid dynamics \cite{fluid1,fluid2},  physical \cite{phys1,phys2} as well as biological systems  \cite{bio1}.   Hence special attention is deemed towards   enriching our knowledge on counter-rotation induced collective dynamics in various systems.   Originally, the coexistence of clockwise and anticlockwise rotations was identified by Tabor \cite{tabor}.  Later, the universal occurrence of mixed synchronization was reported in Stuart-Landau and R\"ossler oscillators \cite{cr1}.  Further, mixed synchronization was demonstrated both theoretically as well as experimentally in the various limit cycle and chaotic systems \cite{cr2,cr3,cr4}. Very recently, the counter-rotating frequency induced dynamical effects were reported in the coupled Stuart-Landau oscillator with symmetry preserving as well as symmetry breaking couplings.   These   authors identified the suppression of oscillation death state when the symmetry breaks in the counter-rotating system \cite{main1_awa}. In addition, feedback also plays an essential role in the various natural and man-made systems. One of the recent reports demonstrated that by introducing a feedback strength in the coupling can switch the stability of the stable steady states facilitating the revival of oscillation in coupled nonlinear oscillators, which was also further corroborated experimentally using coupled electrochemical oscillators \cite{feed_dv1}.   The addition of external feedback in retaining and enhancing the dynamical robustness in the presence and absence of time-delay was also reported \cite{feed2}.     In particular, mean-field feedback has a predominant application in neuroscience in deep brain simulation \cite{neuro1,neuro2}.  

\par Motivated by the above, in this work, we investigate the effect of coexisting co- and counter-rotating  oscillators on the emerging collective dynamical behavior of such an ensemble.  Further, we also aim to study the impact of the mean-field feedback on the observed collective dynamical behaviors.    In order  to elucidate  the above, we   consider  an array of globally coupled Stuart-Landau oscillators. Primarily, the effect of the counter-rotating oscillators will be analyzed in symmetry preserving coupling. Surprisingly, we   find that the proportion of the counter-rotating oscillators play a vital role in facilitating the onset of aging transition through the Hopf bifurcation, which was otherwise identified only due to increasing in the number of inactive oscillators \cite{dai1,dai2,dai3,dai4,bhumi,srilena}. The spread of aging transition is found to increase upon increasing the  frequency of the oscillators. We have also estimated   the critical ratio of the counter-rotating oscillators  to show the stable boundaries of the aging transition. In addition,  we also investigate the effect of  the mean-field feedback on  the  spread of the aging transition.     Interestingly, we find that even a very small decrease in the   mean-field feedback strength  enhances the  aging transition region. In particular, we find that the mean-field induces the phenomenon of aging transition in an ensemble of oscillators even without any counter-rotating oscillators. Further, the robustness of the observed results  will be  analyzed for   symmetry breaking conjugate coupling.  We observe that the  dynamical transition takes place from oscillatory to oscillation death state via aging transition.   We deduce the critical ratio of the  counter rotating oscillator to find the stable boundaries of  the spread of the aging transition. Enriching the observed results, we find that the conjugate coupling, which breaks the rotational symmetry, indeed induces aging transition via  a Hopf bifurcation even without the counter-rotating oscillators and for the unit value of  the mean-field feedback. In this case, the mean-field feedback enhances the aging transition region. It is also to be noted that the transition from aging transition to oscillation death takes place through pitchfork bifurcation.   We  find that the aging transition region enlarges  upon decreasing the feedback strength and increasing the frequency of the oscillators. 
  
	\par The rest of the article is organized as follows:   We will demonstrate the emergence of the  aging transition region through symmetry preserving coupling  in Sec.~\ref{sp}, where the influence of the mean-field feedback will be discussed.       In Sec.~\ref{sbc}, the robustness of the obtained results will be inspected in a symmetry breaking conjugate coupling as well.    Finally, we summarize our results and draw conclusions in Sec.~\ref{concl}.
	
\section{\label{sp} Aging transition: symmetry preserving coupling}
To find the impact of the counter-rotating oscillators in inducing  the aging transition, we consider a general  paradigmatic model of Stuart-Landau limit cycle oscillators with symmetry preserving coupling.  Various natural phenomena including the breathing cycle, the circadian clock shows the limit cycle behavior. Many nonlinear dynamical systems near the Hopf bifurcation can be approximated through  Stuart-Landau oscillator\cite{stu1, stu2, stu3}. An array of globally coupled Stuart-Landau oscillators with symmetry   preserving coupling   can be written as
\begin{eqnarray}
\dot{z}_j = (\lambda+i\omega_j-|z_j|^2)z_j+\frac{K}{N}\sum_{k=1}^{N} (\alpha\,{z_k}-\, {z_j}), 
\label{model_sp}
\end{eqnarray}
where  $z_j=re^{i\theta_j} = x_j+iy_j  \in    C$,  $j=1,2,...,N$.   $N$ be the number of elements in the network which is chosen as $N=100$.    $x_j$ and $y_j$ are the state variables of the $j^{th}$ system.    $\lambda$ is the Hopf bifurcation parameter and $\alpha$ is the mean-field feedback strength.  $\omega_j$ is the  frequency of the $j^{th}$ system. If the system frequency is $+\omega$, the system rotates in a counter-clockwise direction, while $-\omega$ indicates the clockwise direction. In order to   study the role of the counter-rotating oscillators, we split the oscillators as one group of oscillators with the system frequency   $\omega_j=\omega$ for   $j\in {1,...N(1-p)}$   while   the other group takes the value, $\omega_j=-\omega$ for $j\in {N(1-p)+1,...N}$.  The parameter $p$ characterizes the fraction of the counter rotating oscillators. Now, we investigate the emergence of  the aging transition as a function of  the ratio of the counter rotating oscillators.    The numerical analysis of the system~(\ref{model_sp}) is carried out using   Runge-Kutta fourth order scheme with  step  size   0.01 time units.  

Initially, to  unravel the onset of the aging transition as a function of  the ratio of  the counter-rotating oscillators,  we estimate the order parameter $|Z|$,   where $Z=\frac{1}{N} \sum_{j=1}^{N}z_j$, as in refs.~\cite{dai1,dai2,dai3,dai4,bhumi,srilena}. Then the normalized order parameter can be expressed as, $Q\equiv|Z(p)|/|Z(0)|$. $p$  is the ratio  of the counter-rotating oscillators.   $p=0$  and $p=1$,  respectively, denotes the completely    counter-clockwise and clockwise rotating oscillators  while $p=0.5$ indicates the equal ratio of co- and counter-rotating oscillators.  The coupled Stuart-Landau oscillator (\ref{model_sp}), manifests symmetric dynamical  transitions either by increasing the ratio of the counter-rotating oscillators from   the range $p\in (0, 0.5)$  to $p\in (0.5, 1)$  or  decreasing it from  $p\in (1, 0.5)$ to $p\in (0.5, 0)$.  Hence, we examine the effect of the counter-rotating oscillator ratio  from $p=0$ to $p=0.5$ throughout our study.     
\begin{figure}[h!]
	\centering
	\hspace{-0.1cm}
	\includegraphics[width=8.50cm]{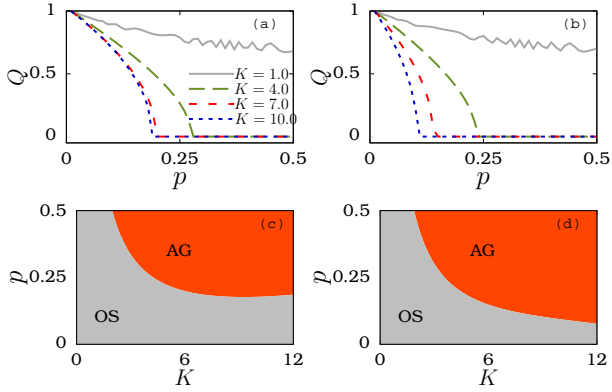}
	\caption{Normalized order parameter ($Q$) as a function of the ratio of the counter-rotating oscillators ($p$)  for  (a)  $\alpha=1.0$  and (b) $\alpha=0.95$,  by fixing   $\omega=5.0$,  $\lambda=1.0$ and $N=100$. The distinct line types denotes the coupling strengths $K=1.0,~4.0,~7.0$ and $K=10.0$.     The corresponding two parameter diagram in $(K,p)$ space  as a function of the coupling strength are depicted in Figs.~(c) and (d).   OS and AG  represent the oscillatory and the aging transition region.}
	\label{fig1}
\end{figure} 

To delineate the emergence of the aging transition, we have plotted the  normalized order parameter (Q) as a function the ratio of the counter-rotating oscillators ($p$) for two different values of the mean-field feedback strength ($\alpha$)  in Figs.~\ref{fig1}(a) and \ref{fig1}(b).  In order to exemplify the above,  we fix the system frequency $\omega=5.0$.  Finite  non-zero value  of   the normalized order parameter is observed   in the entire range of $p$ for $K=1.0$, elucidating the  oscillatory nature of the dynamical state of all the oscillators. It is observed that the normalized order parameter transit from finite value to null value  at the critical values $p_{_{HB}}= 0.28,~0.2,$ and $0.19$,   for   the coupling strengths   $K=4.0,~7.0,$ and  $10.0$,  respectively.   The null value of $Q$ indicates   the existence of the  aging transition region, which emerges through the Hopf bifurcation.  From Fig.~\ref{fig1}(a), it is also  evident that increasing the coupling strength decreases the critical value  $p_{_{HB}}$, thereby illustrating that the aging transition   emerges even for smaller proportion of the counter-rotating oscillators,   enhancing the aging transition region, in strong contrast to increasing the number of inactive oscillators to realize the aging transition as reported so far \cite{dai1,dai2}.  Further, to unravel the influence of the mean-field feedback on the aging transition, we have also depicted the normalized order parameter for $\alpha=0.95$ in Fig.~\ref{fig1}(b).  The coupled systems exhibit  oscillatory behavior for $K=1$.  When increasing the coupling strength to  $K=4.0,~7.0,$ and  $10.0$, the onset of  the aging transition  take place at  $p_c=0.24, ~0.15$ and $0.1$, respectively.  Thus, it is evident from both    Figs.~\ref{fig1}(a)  and \ref{fig1}(b) that the critical value of $p_{_{HB}}$,  for the onset the aging transition,  abruptly decreases even for a feeble decrease in the mean-field feedback strength $\alpha$, enhancing the aging transition region.       The corresponding global behavior  of the coupled co- and counter-rotating  system~(\ref{model_sp})  is depicted in  Figs.~\ref{fig1}(c) and \ref{fig1}(d) in  ($K,p$) space for $\alpha=1.0$ and $\alpha=0.95$, respectively. The co- and  counter-rotating coupled Stuart-Landau oscillators exhibit oscillations for small values of the coupling strength and for the  smaller   ratio of the counter-rotating oscillators, whereas the coupled Stuart-Landau oscillators shows the aging transition   for appreciable values of $p$ and $K$ (see Figs.~\ref{fig1}(c) and \ref{fig1}(d)).    Further, to confirm the emergence of aging transition  analytically, we have performed the linear stability analysis which will be detailed as follows.
\par  In the aging transition, $z_j$ is stabilized to  trivial fixed point (i.e. $z_j=0$). We assume that the coupled system~(\ref{model_sp})  subdivided into two groups, similar to the case of active-inactive group of oscillators  as in   ref.~\cite{dai1}.  Here,    one group  $w_1=z_j, {(j=1,...N(1-p))}$ rotates in    clockwise direction while the  other group $w_2=z_j, (j={N(1-p)+1,...N})$   rotates in  counter-clockwise direction.   

\begin{eqnarray}
\dot{w}_1 &=& (1+i\omega-|w_1|^2)w_1+K(\alpha(1-p)w_2-(1-\alpha p) \,w_1), \nonumber\\
\dot{w}_2&=&(1-i\omega-|w_2|^2)w_2+K(\alpha p w_1+(\alpha(1- p)-1) \,w_2).
\label{ls}
\end{eqnarray}
  It is to be noted that the  sign of $\omega$, either positive or negative depicts whether the oscillators are rotating in clockwise or anti-clockwise direction, respectively.   By performing the  linear stability analysis of Eq.~(\ref{ls}) around the origin ($w_1 = w_2=0$), one can obtain an expression for the critical ratio of the counter rotating oscillators $p_{_{HB}}$ as 
\begin{eqnarray}
p_{_{HB}} =\frac{\alpha^2 K^2 (2 + (-2 + \alpha) K)^2 \omega^2 \pm \sqrt{\Delta_1}}{(2 \alpha^2 K^2 (2 + (-2 + \alpha) K)^2 \omega^2)}.  
\label{sc2}
\end{eqnarray}
where, $\Delta_1=\alpha^2 K^2 (2 + (-2 + \alpha) K)^6 \omega^2 (-1 - (-2 + \alpha) k + (-1 + \alpha) K^2 - \omega^2)$. 
\begin{figure}[h!]
	\centering
	\hspace{-0.4cm}
	\includegraphics[width=8.5cm]{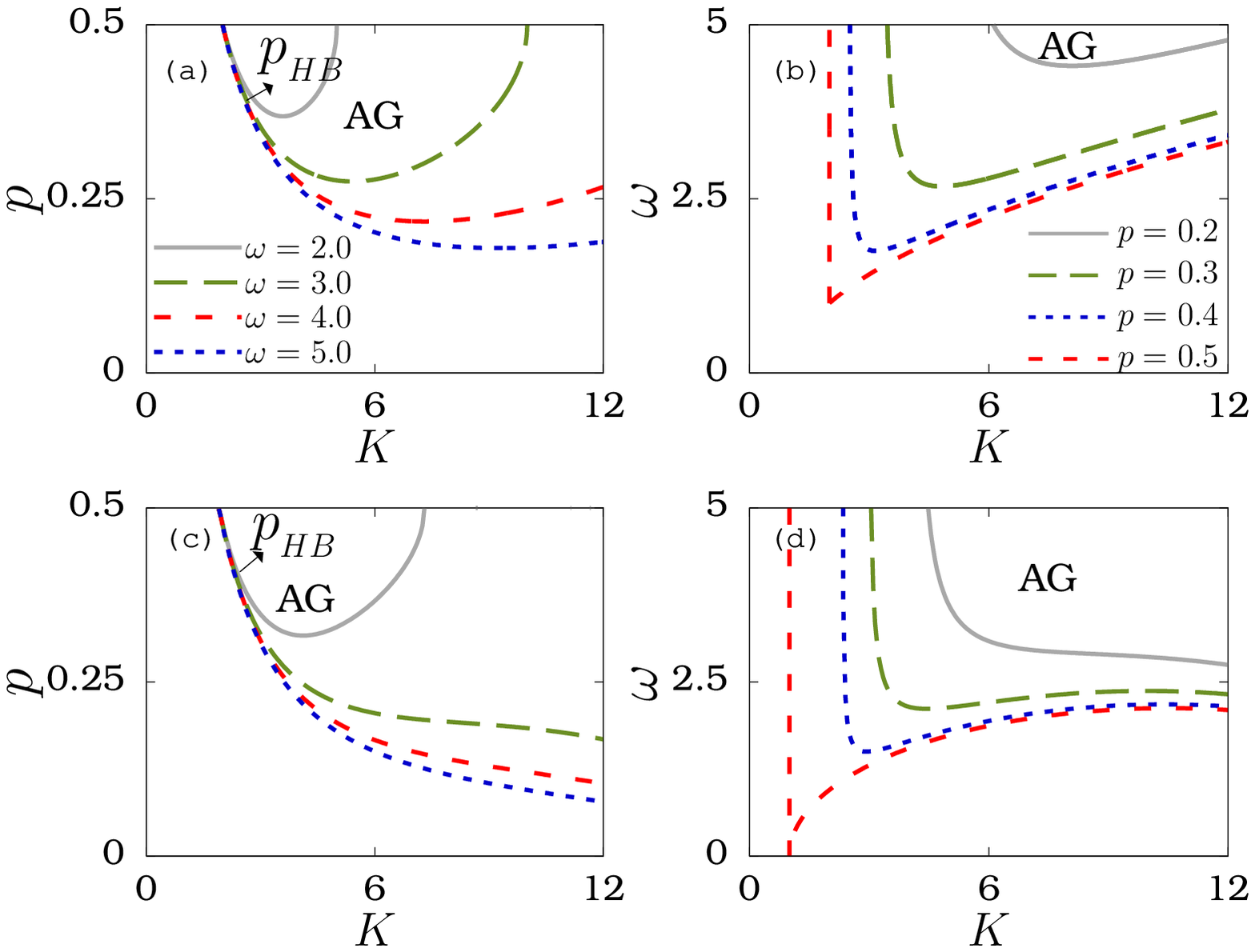}
	\caption{Analytical boundaries of  the aging transition region for $\alpha=1.0$: (a) $(K,p)$ space  for different values of system frequency $\omega=2.0,~3.0,~4.0$ and $5.0$, (b) $(K,\omega)$ space  for different values of the critical ratio of the counter rotating oscillators $p$ for $p=0.2, 0.3,  0.4 $ and $p=0.5$. {  $p_{_{HB}}$  is the  critical  curve corresponding to Hopf bifurcation curve. } The corresponding analytical aging boundaries for $\alpha=0.95$  are shown in Fig.~(c) and (d).}
	\label{fig2}
\end{figure}
Using the above expression, one can obtain the stable boundaries of the aging transition region as  depicted  in  Fig.~\ref{fig2}.  {  The critical curve $p_{_{HB}}$ is the Hopf bifurcation curve across  which there is a change the stability of the limit cycle occurs.  }  Figures~\ref{fig2}(a) and \ref{fig2}(b) are  plotted  for the mean-field feedback $\alpha=1.0$.      The distinct  line-types in Fig.~\ref{fig2}(a) correspond to the different values of $\omega$, namely as  $\omega=2.0$, $\omega=3.0$, $\omega=4.0$ and $\omega=5.0$  in $(K, p)$ plane. From Fig.~\ref{fig2}(a), it is evident that increasing the system frequency increases the aging transition region as a function of the coupling strength and the ratio of the counter-rotating oscillators. Similarly, the aging transition regions are depicted in $(K,  \omega)$ plane in Figure~\ref{fig2}(b) for different   ratio of the counter rotating oscillators   $p=0.2,~p=0.3, ~p=0.4 $ and $p=0.5$. The region enclosed by the curves correspond to aging transition region.  Figure~\ref{fig2}(b) clearly delineates the increase in the    aging transition region while increasing the ratio of counter-rotating oscillators.  Moreover, the aging transition region enhances to a large extent when both the co- and counter-rotating oscillators balance each other.  Further, to substantiate the observed results in Fig.~\ref{fig1}, we have also depicted the  dynamical transition observed for the mean-field feedback $\alpha=0.95$.   Thus,  it is evident from Figs.~\ref{fig2}(c) and \ref{fig2}(d)  that the aging transition region is increased upon decreasing the mean-field feedback strength $\alpha$.   In particular, the aging transition region enhances for increasing the  frequency of the oscillations, increasing the ratio of the  counter-rotating oscillators and while decreasing the mean-field feedback strength. 
\begin{figure}[ht!]
	\centering
	\hspace{-0.3cm}
	\includegraphics[width=8.50cm]{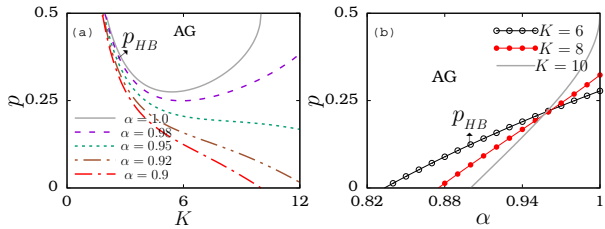}
	\caption{(a) The aging transition  region, for symmetry preserving coupling, for different values of  the mean-field  feedback strength ($\alpha=1.0,~ 0.98,~ 0.95,~ 0.92$ and $0.9$)  for the system frequency $\omega=3.0$.  (b) Critical ratio of counter rotating oscillators as a function of the feedback strength for the system frequency $\omega=3.0$ and the coupling strengths $K=6.0$,  $K=8.0$  and $K=10.0$.}
	\label{fig3}	
\end{figure} 
\par For  further insight on  the effect of  the mean-field feedback strength on the aging transition region   and to show the enlarging of the aging transition region as a function of the feedback strength, we have depicted the critical boundaries  $p_{_{HB}}$ of the aging transition for $\omega=3.0$ in Fig.~\ref{fig3}(a).  The boundaries are plotted for the values of the  feedback strength $\alpha=1.0,~0.97,~0.95,~0.92$ and $0.9$ which clearly depicts the   increase in the aging transition region   for decreasing  values of the  feedback strength. Further, the observed results are validated by finding the critical value   $p_c$ as a function of $\alpha$ for   three different values of the coupling strengths $K=6.0$,  $K=8.0$  and $K=10.0$ as shown in Fig.~\ref{fig3}(b). Lines connected by open circles, filled circles and solid line correspond to the coupling strength  $K=6.0$,  $K=8.0$  and $K=10.0$, respectively.     It is evident that the critical value $p_c$ facilitating the onset of the aging transition region drops down to zero  even for a small decrease in the mean-field feedback strengths.    Substantiating that the aging transition can indeed induced by the mean-field feedback strength even in the absence of  the counter-rotating oscillators.  One can also note that increasing the coupling strength decreases the value of   the  mean-field feedback strength for the onset of the aging transition.

\par   From symmetry preserving coupling, we have  unraveled   the emergence of the aging transition via Hopf bifurcation as a result of increasing the ratio of the counter-rotating oscillators. It is also established that the mean-field feedback can facilitate the onset of the aging transition without any counter-rotating oscillators.  In addition, the spread of the aging transition region is increased upon either increasing the system frequency or decreasing the mean-field feedback strength. 
Further, to examine the robustness of the  observed  transition,   we have also analyzed the dynamical transitions in  a  symmetry breaking coupling in the following. 
\section{\label{sbc}Aging transition: symmetry  breaking coupling}   
It is established fact that the symmetry breaking coupling can  be responsible for the emergence of various collective patterns including frequency cluster, amplitude cluster, amplitude chimera, frequency chimera, various kinds of oscillation quenching states and so on \cite{sb1,conj1,prema1}. We introduce the symmetry breaking in the coupling using the conjugate variables in the coupling. In this section, we investigate  the effect of conjugate coupling on the aging transition as a function of the  counter-rotating oscillators, the frequency of the oscillators, and the mean-field feedback.  In order to illustrate the above,  we consider an array of globally coupled Stuart-Landau oscillators  with conjugate coupling described as
\begin{eqnarray}
\dot{z_j} = (\lambda+i\omega_j-|z_j|^2)z_j+\frac{K}{N}\sum_{k=1}^{N}  (\alpha\, Im({z_k})- Re({z_j})).
\label{model_sb}
\end{eqnarray}
Here, the coupling among the oscillators are implemented through   dissimilar variables.  The proposed coupling breaks the system rotational symmetry explicitly, where, the rotational invariance is not preserved under such transformation $z_j\rightarrow z_j e^{i\theta}$. 
\begin{figure}[ht!] 
	\centering
	\hspace{-0.1cm}
	\includegraphics[width=8.50cm]{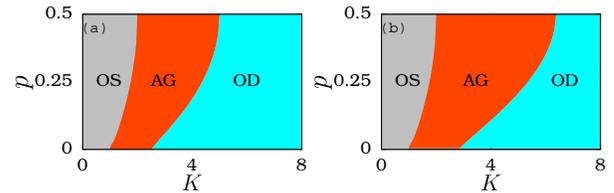}
	\caption{Two parameter diagram for  an array of conjugately coupled  Stuart-Landau oscillators (symmetry breaking coupling) in $(K,p)$ space for    (a) $\alpha=1.0$ and (b) $\alpha=0.96$.  Other parameters: $\lambda=1.0$, $\omega=2.0$ and $N=100$. OS, AG and OD represent the oscillatory, aging transition state and oscillation death state, respectively.  }
	\label{fig4}
\end{figure} 
\par Initially, the dynamical transitions are analyzed numerically as a function of the coupling strength in symmetry breaking conjugately coupled system~(\ref{model_sb}).  Figures~\ref{fig4}(a) and \ref{fig4}(b) are plotted for two different values of the mean-field feedback strength $\alpha=1.0$ and $\alpha=0.96$, respectively, for the system frequency  $\omega=2.0$.  From both the figures, we found that the transition takes place  from   oscillatory  region to oscillation death state through the aging transition. The transition from oscillatory to aging transition transition takes place via Hopf bifurcation while transit from aging to oscillation death occurs via pitchfork bifurcation. Further, it is to be noted that the symmetry breaking coupling using the conjugate variables induces  the aging  transition without any counter-rotating oscillators even for the unit value of the mean-field feedback strength.   Comparing Fig.~\ref{fig4}(b) with Fig.~\ref{fig4}(a), it is clear that  the feedback strength  $\alpha$   increases the aging transition region in the $(K,p)$ space but the oscillatory region remains unaltered. Furthermore, the observed numerical boundaries of the aging  transition region are confirmed analytically   in the following.

\begin{figure}[ht!] 
	\centering
	\hspace{-0.1cm}
	\includegraphics[width=8.50cm]{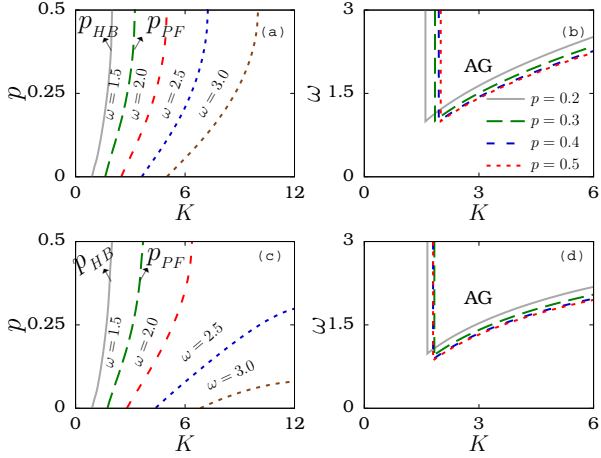}
	\caption{Analytical boundaries of the aging transition region      (a) ($K,p$) plane for distinct values of the  system frequency $\omega=2.0,~3.0,~4.0$ and $5.0$ and  (b)     $(K, \omega)$ plane for  $p=0.2, 0.3,  0.4 $ and $p=0.5$.  $p_{_{HB}}$ and $p_{_{PF}}$ are the critical value of the counter rotating oscillators, which separate the boundary between the oscillatory and the aging transition region, and the aging transition and the  oscillation death, respectively. The corresponding  analytical boundaries for $\alpha=0.96$  is shown in Fig.~(c) and (d). }
	\label{fig5}
\end{figure} 

\par For finding the   analytical boundaries of the aging transition region in the symmetry breaking conjugate coupling, we  have deduced the  critical ratio of the counter rotating oscillators from the linear stability analysis as, \\
\begin{eqnarray}
p_{_{HB}}=  \frac{(2 \alpha^2 (-1 + K)^2 K^2 w^2 \pm \sqrt{\Delta_2})}{4 \alpha^2 (-1 + K)^2 K^2 w^2},
\label{eq1}
\end{eqnarray}
and
\begin{eqnarray}
p_{_{PF}}=  \frac{\alpha^2K^2\omega^2 \pm \sqrt{\Delta_3})}{2\alpha^2 K^2 w^2},
\label{eq2}
\end{eqnarray}
where  $\Delta_2= \alpha^2 (-1 + k)^4 K^2 (4 - 
8 K - (-4 + \alpha^2) K^2) \omega^2 (8 K + (-4 + \alpha^2) K^2 - 4 (1 + \omega^2))$, and  $\Delta_3=\alpha^2K^2\omega^2((\alpha^2-1)K^4-2(\alpha^2-2)K^3+(\alpha^2-6-2\omega^2)+4K(1+\omega^2)-(1+\omega^2)^2)$.    $p_{_{HB}}$ and $p_{_{PF}}$  are the critical curves corresponding to the  Hopf bifurcation and the pitchfork bifurcation curves. Across   $p_{_{HB}}$ there is a change in the stability of the limit cycle oscillators, which separate the boundary between the oscillatory state and the aging transition region.    $p_{_{PF}}$  is the pitchfork bifurcation curve on which there is a transition from homogeneous steady state (AT) to inhomogeneous steady state (OD state).   Using Eq.~(\ref{eq1}) and Eq.~(\ref{eq2}), we have plotted  Fig.~\ref{fig5} in ($K,p$) space and ($K,~\omega$) space. Enhancing of the aging transition region is clearly evident (see  Fig.~\ref{fig5}(c)  and  \ref{fig5}(d))  for decrease in the value of the mean-field strength.  

\par
\begin{figure}[ht!]
	\centering
	\hspace{-0.1cm}
	\includegraphics[width=9.00cm]{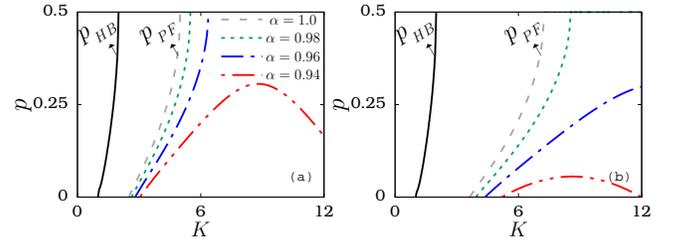}
	\caption{The  aging transition region for the different value of  the feedback strength in conjugately coupled Stuart-Landau oscillators. Analytical boundaries of the aging transition region for distinct values  of feedback strength $(\alpha=1.0,~0.98,~0.96,~0.94)$ for (a)  $\omega=2.0$ and  (b) $\omega=2.5$.   }
	\label{fig6}
\end{figure}

 Finally, to show  the nature of the aging transition region upon decreasing the mean-field feedback strength  in the symmetry breaking coupling, we have depicted the analytical curves of different  mean-field feedback strength $\alpha=1.0,~0.98,~0.96,~0.94$  for  $\omega=2.0$ and $\omega=2.5$ in Figs.~\ref{fig6}. It is evident from Figs.~\ref{fig6}(a) and \ref{fig6}(b)  that the aging transition region increases upon increasing the frequency of the oscillators  and decreasing the mean-field feedback strength.  Further, it is also evident that decreasing the feedback strength to $(\alpha=1.0,~0.98,~0.96,~0.94)$ increases the aging transition region. 
\par  From the above discussed results from the  symmetry preserving and symmetry breaking coupling, it is confirmed  that  even the proportion of the counter-rotating oscillator in an ensemble of coexisting co- and counter-rotating oscillators, indeed, facilitates the emergence of the aging transition in  contrast to the earlier reports, where a large proportion of inactive oscillators are introduced to realize the same.  In addition, it is shown that the mean-field feedback strength facilitates  the onset of the aging transition despite the absence of the counter-rotating oscillators.  Region of the observed aging  transition is enhanced while increasing the frequency of the system or decreasing the mean-field feedback strength.   Further, it is clear that such transition is robust in both  symmetry preserving as well symmetry breaking coupled oscillators.

\section{\label{concl}Conclusion}	
Degradation of dynamical activity (is also called as aging) is inevitable in many complex networks including power grids and neural networks.  In the present work, we have explored the onset of the aging transition  in a network of the globally coupled limit cycle oscillators where the system frequencies are distributed to a distinct ratio of  co- and counter-rotating oscillators.  We found that  even the counter-rotating oscillators also can induce the   aging transition in an ensemble of coexisting co- and counter-rotating oscillators unlike the  earlier reports, which increases a large proportion of the inactive oscillators. In order to show this, initially, we have considered an array of globally coupled Stuart-Landau limit cycle oscillators with symmetry preserving coupling. We have analyzed the  influence of the mean-field feedback strength  on the aging transition.   Interestingly,   we have observed that    aging transition region is enhanced abruptly even for a small decrease in the feedback strength and subsequently the ratio of the counter-rotating oscillators necessary to induce the aging transition drops down to zero corroborating that the mean-field feedback can indeed induce  the phenomenon of  the aging transition.   Further,  we have also analyzed the effect of the system frequency on the aging transition region, which increases the aging transition region.  The observed results are further confirmed analytically through the linear stability analysis by deducing the critical ratio of the counter-rotating oscillators for which change in the stability of the limit cycle oscillators occurs via the Hopf bifurcation.   Additionally,   the robustness of such aging transition is also analyzed in the symmetry breaking conjugate coupling which also exhibits similar dynamical transitions. Surprisingly, we found that the symmetry breaking conjugate coupling itself is capable of facilitating the onset of the aging transition without any counter-rotating oscillators even for the unit value of the mean-field feedback strength.   Furthermore, we have also deduced the  analytical critical curves corresponding to the Hopf bifurcation and the pitchfork bifurcation curves.  Finally, from the obtained results, we have identified that the observed aging transition behavior is robust in both symmetry preserving as well as symmetry breaking coupling. \\ 

\par The proposed work also leads to many open problems. Due to complex interactions in  real-world systems, the study of collective dynamics under distinct network geometries is intriguing research for many years and is inevitable.  Therefore, it is a valuable insight to extend our analysis to various topological structures including scale-free, small world and so on \cite{dai2,srilena}. Since, feedback is a general mechanism to revoke the dynamism from the deteriorated dynamical units in a complex network and is also essential to investigate the influence of various feedbacks \cite{feed_dv1,feed1,feed1_vkc,feed_pon}.    From the earlier reports, it is identified that the processing delay is used to destabilize the homogeneous as well as an inhomogeneous steady state either in the presence or the absence   of propagation delay  \cite{pro_delay,dv_2019}.   Thus, it will also be interesting to investigate the effect of various delays including propagation delay, processing delay, and low pass filter \cite{feed1_vkc,feed_dv1,pro_delay,feed_pon,dv_2019,lpf1}. Furthermore, the role of nonlinear parameter is inevident in the observed dynamical transitions and which plays a crucial role in inducing various collective patterns \cite{dai2,dai3,sb1,prema1}.  In view of the above, it is also intriguing to investigate the dynamical transition with the addition of nonisochronicity parameter, as a function of co-  and counter-rotating frequencies of a dynamical system.
\begin{acknowledgments}
KS sincerely thanks the CSIR for a fellowship under SRF Scheme (09/1095(0037)/18-EMR-I). The work of VKC supported by a research project   CSIR  under Grant No.03(14444)/18/EMR-II.  DVS is supported by the CSIR EMR Grant No. 03(1400)/17/EMR-II. We also thank the SASTRA Deemed university for providing good infrastructure lab facilities.\\  
\end{acknowledgments} 

\end{document}